\documentstyle[epsfig,aps,amssymb]{revtex}

\newcommand{\gb}{\beta}

\newcommand{\gk}{\kappa}

\newcommand{\gl}{\lambda}

\newcommand{\gD}{\Delta}

\tightenlines
\begin{document}


\twocolumn[
\hsize\textwidth\columnwidth\hsize\csname @twocolumnfalse\endcsname


\title{Three-quark clusters at finite temperatures and densities}
\author{M. Beyer\footnote{corresponding author: Fachbereich Physik,
    Universit\"at Rostock, 18051 Rostock,
    Germany, email:michael.beyer@physik.uni-rostock.de,
    Tel.:[+49](0)381/4981674, Fax.:[+49](0)381/4981673}, S. Mattiello}
\address{Fachbereich Physik, Universit\"at Rostock, 18051 Rostock,
  Germany} \author{T. Frederico} \address{Dep. de F\'\i sica,
  Instituto Tecnol\'ogico de Aeron\'autica,
  Centro T\'ecnico Aeroespacial, \\
  12.228-900 S\~ao Jos\'e dos Campos, S\~ao Paulo, Brazil.}
\author{H.J. Weber} \address{Dept. of Physics, University of Virginia,
  Charlottesville, VA 22904, U.S.A.}

\date{\today}
\maketitle
\begin{abstract}
  We present a relativistic three-body equation to study correlations
  in a medium of finite temperatures and densities. This equation is
  derived within a systematic Dyson equation approach and includes the
  dominant medium effects due to Pauli blocking and self energy
  corrections. Relativity is implemented utilizing the light front
  form. The equation is solved for a zero-range force for parameters
  close to the confinement-deconfinement transition of QCD.  We
  present correlations between two- and three-particle binding
  energies and calculate the three-body Mott transition.
\end{abstract}

\vspace{5mm}

\vspace{5mm}
\vspace{5mm}

]

Exploring the phase structure of quantum chromodynamics (QCD) for the
whole density-temperature plane is a challenging task for the standard
model.  Lattice Monte Carlo simulations have revealed several exciting
results over the past decade~\cite{Karsch:2000vy,Alford:1999sd}. In
addition, at finite densities, modeling of QCD has added substantially
to our understanding of its rich phase structure.

One particularly interesting result is the possibility of
``color-superconductivity'', extensively discussed
recently~\cite{Alford:2001dt,Alford:1998zt,Rapp:1998zu,Blaschke}. The
possible appearance of color-superconductivity is related to similar
ideas that led to nucleon or Cooper pairing.  Another important aspect
is the transition from nucleons to quarks as relevant degrees of
freedom (Mott transition) that is likely to be related to the
``confinement-deconfinement'' phase transition. In these cases the
influence of three-quark correlations has hardly been investigated.
Recently, Pepin et al. addressed the question of a possible
competition of three-quark clusters and two-quark
condensation~\cite{Pepin:2000hs} at finite density but zero
temperature.

The problem of correlations in medium is tackled within a Green
functions formalism~\cite{fet71} using the Dyson equation
approach~\cite{duk98}. Three-particle correlations on the basis of a
new in-medium Alt-Grassberger-Sandhas~\cite{Alt:1967fx} (AGS) type
equation have been investigated in
\cite{Beyer:1996rx,Beyer:1997sf,Beyer:1999tm,Beyer:1999zx,Kuhrts:2000jz}
and implemented into a Boltzmann-Uehling-Uhlenbeck simulation of heavy
ion collisions and compared to experiments
in~\cite{Beyer:2000fr,Beyer:1999xv,Kuhrts:2001zs}.  Recently, the
formalism has been extended to four-particle correlations and solved
to study in-medium properties of the
$\alpha$-particle~\cite{Beyer:2000ds}.  In contrast to the nucleonic
phase, however, investigations of the quark phase of QCD entail
several new issues such as relativity, color degrees of freedom,
confinement, effective residual interactions, number of flavors, among
others.
  
In order to derive effective few-body equations for quarks at finite
temperatures and densities we have to consider in-medium effects as
well as relativity.  In medium the mass of the constituents may vary
and in fact becomes close to zero for higher temperatures and
densities, see e.g.~\cite{Klevansky:1992qe}.  To implement relativity
we use the light front approach~\cite{Dirac:1949cp}. In the context of
the quantum statistical framework it has some compelling advantages,
although the concept of quasi particles in a background field that
will be utilized further down and extended to include correlations
introduces a special frame of reference. These are: i) Several boosts
are kinematical (interaction free).  As a consequence of the
transitivity of kinematical boosts the Fock state decomposition is
stable~\cite{Perry:1990mz}. Since we decouple the hierarchy of Green
function equations using a cluster mean field expansion~\cite{duk98}
this decoupling of Fock spaces is therefore retained. ii) The
dependence of the equations for the isolated system on the c.m.
momentum of the cluster can easily be separated, see
e.g.~\cite{Terentev:1976jk,Bakker:1979eg,Kondratyuk:1980gj}. In a
homogeneous medium this is also an important feature of the light
front form, because inclusion of Pauli blocking factors that depend on
the c.m. momentum of the cluster leads to a parametric dependence on
the c.m. momentum only. iii) Pair creation processes are likely to be
suppressed on the light front~\cite{Frankfurt:1979vc,Frankfurt:1981mk}
in some frames of reference.  Therefore we may presently consider
particles only and no antiparticles which leads to technical
simplifications that, however, can be relaxed as the investigation
goes on. iv) Another advantage is that the light front form can be
formulated in a Hamiltonian language and therefore all the results of
the previously tested quantum statistical framework used here can be
recovered using a proper interpretation.

On the other hand the vacuum structure in the light front dynamics
involves technical difficulties in the presence of condensates, viz.
zero modes in general, see e.g.~\cite{lenz2000}. These difficulties
include a proper description of zero modes and Goldstone modes which
are being investigated by several groups, see e.g.~\cite{pauli2000}.
Further, rotational invariance on the light front implies interaction
dependent generators. As a consequence angular decomposition is
technically more involved. Presently, we make use of angular
averaging. This introduces uncertainties, but for homogeneous infinite
matter angular averaging has proven quite useful in the past.
Finally, the description of spin degrees of freedom on the light-front
is an issue that involves real complications for the three-body
system. For an extensive discussion of the problem see
Ref.~\cite{Beyer:1998xy}.  Presently, we treat the spin in a
simplified manner; namely by averaging over elementary spin
projections. This approximation leads to an effective one-channel
bose-type three-body equation.  This simplifies the calculation
significantly for the sake of a transparent treatment of in-medium
effects of the relativistic equations. A bose-type relativistic
three-body equation with a zero range interaction has previously
proven useful for studies of isolated nucleon form
factors~\cite{Frederico:1992uw,deAraujo:1995mh}. The technically
involved full inclusion of the spin degrees of freedom along the lines
of Ref.~\cite{Beyer:1998xy} will be postponed to a later stage of the
calculations. As a consequence we are presently not able to
distinguish between so-called scalar, pseudoscalar, vector, and axial
vector quark pairs.  As many aspects can be improved, we emphasize
that we present and solve for the first time an in-medium relativistic
three-body equation for both finite temperatures and finite densities.

The light front has been utilized earlier for infinite nuclear matter
calculations at zero temperatures and normal nuclear matter densities
in~\cite{jerry}. That approach is based on the Br\"uckner G matrix and
two-body correlations are treated with an in-medium
Blankenbecler-Sugar equation including a medium-modified interaction.
For a general discussion of differences between the Green function
approach (for finite temperatures and densities) used here and the
Br\"uckner approach (at zero temperatures) see e.g. Ref.~\cite{duk98}.

The equations derived here are based on a systematic quantum
statistical framework using a cluster expansion for the Green functions. 
 We start with the chronological Green
function~\cite{fet71,duk98}
\begin{eqnarray}
i{\cal G}^{\tau-\tau'}_{\alpha\beta} &= &\langle T_\tau A_\alpha(\tau)
A^\dagger_\beta(\tau')\rangle\nonumber\\
&\equiv&\theta(\tau-\tau')\langle A_\alpha(\tau)
A^\dagger_\beta(\tau')\rangle\nonumber\\&& \mp
\theta(\tau'-\tau)\langle  A^\dagger_\beta(\tau')
A_\alpha(\tau)\rangle
\label{eqn:green}
\end{eqnarray}
where $\tau$ denotes the light cone time $x^+=t+z$.  The upper (lower)
sign stands for Fermions (Bosons), $A_\alpha(\tau)$ are arbitrary
Heisenberg operators. For the three-particle problem in question they
are given by Dirac operators taken at equal $\tau$, viz.
$A_\alpha(\tau)=\psi_1(\tau)\psi_2(\tau)\psi_3(\tau)$; indices
represent all other quantum numbers. Note that on the light front the
chronological Green function for spin 1/2 (in this case
$A_\alpha(\tau)=\psi_1(\tau)$) differs from the Feynman propagator by
a contact term. Presently we neglect this term being of higher
order~\cite{Sales:2001gk,deMelo:1999gn}. The generalization to
imaginary time based on the Matsubara techniques is direct and has
been used, e.g., to treat few-body correlations in nucleonic
systems~(see
e.g.~\cite{duk98,Beyer:1996rx,Beyer:1997sf,Beyer:1999tm,Beyer:1999zx,Kuhrts:2000jz,Beyer:2000fr,Beyer:1999xv,Kuhrts:2001zs,Beyer:2000ds}).
The equation of motion (Dyson equation) is given by~\cite{duk98}
\begin{eqnarray}
i\frac{\partial}{\partial\tau} {\cal G}^{\tau-\tau'}_{\alpha\beta}
&=& \delta(\tau-\tau'){\cal N}_{\alpha\beta}+
\sum_\gamma\int d\tau_1 {\cal M}_{\alpha\gamma}^{\tau-\tau_1}
{\cal G}^{\tau_1-\tau'}_{\gamma\beta}
\label{eqn:dyson}
\end{eqnarray}
where
\begin{equation}
{\cal N}_{\alpha\beta} = 
\langle[A_\alpha(\tau),A^\dagger_\beta(\tau)]_\pm\rangle.
\label{eqn:N}
\end{equation}
Here, we neglect retardation in the mass operator ${\cal M}$, that
would introduce more intermediate Fock space components, viz.
\begin{eqnarray}
 {\cal M}_{\alpha\beta}^{\tau-\tau'}
&\rightarrow & \delta(\tau-\tau'){\cal M}^\tau_{0,\alpha\beta},\\
{\cal M}^\tau_{0,\alpha\beta}&=&\sum_\gamma
\langle[[A_\alpha,H](\tau),A^\dagger_\gamma(\tau)]_\pm\rangle
{\cal N}^{-1}_{\gamma\beta}.
\label{eqn:M}
\end{eqnarray}
So far the formalism presented is rather general and for a typical
many-body Hamiltonian (with generic two-body interactions $V_2$) has
been proven useful for various domains of many-body physics
(see~\cite{duk98}) including the calculation of correlations in
nuclear systems at finite densities and
temperatures~\cite{Beyer:1996rx,Beyer:1997sf,Beyer:1999tm,Beyer:1999zx,Kuhrts:2000jz,Beyer:2000fr,Beyer:1999xv,Kuhrts:2001zs,Beyer:2000ds}.

At chemical potentials $\mu$ and temperatures $T=1/(k_B\gb)$ averaging
is due to a grand canonical ensemble in equilibrium, $\langle
O\rangle={\rm Tr}\{\rho_0 O\}$, where $\rho_0$ denotes the
corresponding statistical operator expressed in terms of light front
operators. In averaging we consider the few-body cluster embedded in a
homogeneous mean field of uncorrelated particles which is a reasonable
approximation. Hence, together with (\ref{eqn:M}) the hierarchy of
Green functions decouples.  This leads, upon introducing particle
$b(k)$ and antiparticle $d(k)$ Fock operators, to the standard
Fermi-Dirac single particle distribution functions. For particles this
reads
\begin{eqnarray}
 f(k^+,\vec k^2_\perp)&= &\langle b^\dagger(k)b(k)\rangle
\nonumber\\
&=&\left(\exp\left[\gb\left(\frac{1}{2}(
k^++k_{\rm on}^-)-\mu\right)\right]+1\right)^{-1},
\label{eqn:Fermi}
\end{eqnarray}
expressed in terms of light front form momenta given by $\vec
k_\perp=(k_x,k_y)$ and $k^\pm=k_0\pm k_z$. The on-shell light-front
energy is given by $k_{\rm on}^-=(\vec k_\perp^2+m(\mu,T)^2)/k^+$ with
a medium dependent mass $m(\mu,T)$. The components of the four vector
$k_{\rm on}$ are $k_{\rm on}=(k_{\rm on}^-,\vec k_\perp,k^+)$. The
thermodynamic Green function for the one-particle case, by use of
eqs.~(\ref{eqn:green}), (\ref{eqn:N}), and (\ref{eqn:M}) is then given
by
\begin{eqnarray}
G(k) &=&
\frac{\theta(k^+)}{2k^+}(\gamma k_{\rm on}+m(\mu,T))
\nonumber\\
&&\times\left(\frac{1-f}{\frac{1}{2}k^--\frac{1}{2}k^-_{\rm on}+i\varepsilon}
+\frac{f}{\frac{1}{2}k^--\frac{1}{2}k^-_{\rm on}-i\varepsilon}
\right).
\end{eqnarray}

Presently we consider only particle degrees of freedom on the light
front and have therefore dropped the term related to $\theta(-k^+)$.
The energy variable $k^-$ plays the r\^ole of the Matsubara
frequency~\cite{fet71},
$\frac{1}{2}(k^-+k^+)=k^0\rightarrow\pi\lambda/(-i\beta)+\mu$.
Eventually, averaging over the elementary spin projections leads to a
bose type Green function.

For particles that are part of a larger $n$-body cluster it is
convenient to introduce fractions $x=k^+/P_n^+$, where $P^+_n$ is the
plus component of the cluster's c.m. momentum. For a cluster at rest
$P^+_n=M_n$, where $M_n$ is the mass of the cluster.

\begin{figure}
\begin{center}
\epsfig{figure=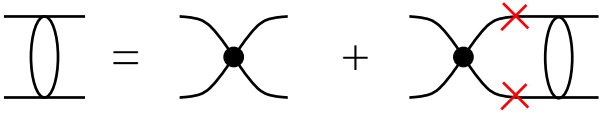,width=0.45\textwidth}
\end{center}
\caption{\label{fig:T2}
  Feynman-Galitzkii equation for the two-body $t$-matrix with zero
  range interaction. The crosses refer to the Pauli-blocking factor
  $N_2$. Lines represent quasi-particles.}
\end{figure}
Evaluation of Eq.~(\ref{eqn:dyson}) for two or three particles with
the mass operator of eq.~(\ref{eqn:M}) in a homogeneous medium of
independent particles leads to resolvent equations of the Green
function where the single time $\tau-\tau'$ is Fourier transformed to
a Matsubara frequency. These equations include mass corrections and
Pauli blocking factors in a systematic way. They can be rearranged
into equations for $t$-matrices. The resulting equation for the
two-body $t$-matrix $T_2$ has the same formal structure as the
Feynman-Galitzkii equation shown in Fig.~\ref{fig:T2}
\begin{equation}
T_2=V_2 + V_2 R_0 N_2  T_2
\label{eqn:T2}
\end{equation}
where $V_2$ represents the two-body interaction, and $R_0$ the
interaction independent two-body resolvent.  The Pauli blocking factor
$N_2$, represented by the crosses in Fig.~\ref{fig:T2}, is given by
\begin{equation}
N_2 = \bar f_1 \bar f_2 - f_1 f_2,\qquad \bar f=1-f,
\end{equation}
where the indices of the Fermi-Dirac function $f$ reflect particle
quantum numbers.

\begin{figure}
\begin{center}
    \epsfig{figure=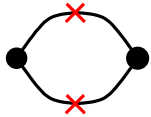,width=0.11\textwidth}
\end{center}
\caption{\label{fig:loop}
  Loop diagram corresponding to the kernel of the integral equation
  (\protect\ref{eqn:tau}).  The crosses refer to the Pauli-blocking factor
  $N_2$.}
\end{figure}
For a numerical analysis and a first calculation of in-medium effects
we utilize a zero-range model studied earlier in a different
context~\cite{Frederico:1992uw,deAraujo:1995mh}. The equation
represented by Fig.~\ref{fig:T2} can be summed and leads to a solution
for the two-particle propagator $\tau(M_2)$, i.e.
\begin{equation}
\tau(M_2)=\left(i\gl^{-1} - B(M_2)\right)^{-1}.
\label{eqn:tau}
\end{equation}
The expression for $B(M_2)$ is represented by the loop diagram of
Fig.~\ref{fig:loop} and, in the rest system of the two-body system
$P^\mu=(M_2,0,0,0)$, given by
\begin{equation}
B(M_2)=-\frac{i}{(2\pi)^3} \int \frac{dx d^2k_\perp}{x(1-x)}
\frac{1-F(x,\vec k_\perp^2)}{M_2^2-M_{20}^2},
\label{eqn:B}
\end{equation}
where
\begin{eqnarray}
M_{20}^2&=&(\vec k_\perp^2+m^2)/x(1-x),\\
F(x,\vec k_\perp^2)&=&f(x,\vec k^2_{\perp})+f(1-x,\vec k_\perp^2)
\end{eqnarray}
and $f$ given in eq.~(\ref{eqn:Fermi}) with $x=k^+/P^+_2$.
The integral can obviously be separated into a blocking independent
term $B_0(M_2)$ and the change $\Delta B(M_2)$ depending on the Fermi
functions $F(x,\vec k_\perp^2)$. The integral involving $B_0(M_2)$ has a
logarithmic divergence that can be absorbed in a redefinition of
$\gl$. The physical information introduced in the renormalization of
the amplitude is the mass of the two bound particles, $M_{2B}(\mu,T)$.
If we assume that the two particle amplitude $\tau$ has a pole
for $M_2(\mu,T)=M_{2B}(\mu,T)$ we may write
\begin{equation}
i\gl^{-1}=B(M_{2B})\equiv B_0(M_{2B}) + \gD  B(M_{2B}),
\label{eqn:Bound}
\end{equation}
where the dependence on $T$ and $\mu$ is suppressed in the notation.
Although, we use eq.~(\ref{eqn:Bound}) for finite temperatures and
densities, in principle, it is enough to assume such a bound state for
the isolated case only, and calculate the corresponding in-medium
bound state $M_{2B}(\mu,T)$ from the above equations. This would
require a definite regularization procedure which we presently want to
omit. The assumption of a bound state implies no restriction on
the conclusions of our investigation concerning the relative
importance of two- and three-body correlations at a given temperature
and density.  This will be obvious from the results presented and
discussed at the end of this Letter.

For $M_2(\vec P_2=0,\mu,T)<2\mu$ (for $\mu<m$) and $2m<M_2(\vec
P_2=0,\mu,T)<2\mu$ (for $\mu>m$) instabilities of the Fermi gas
against formation of Bose condensation or Cooper pairing can occur. As
a consequence the distribution functions need to be modified to
include the gap energy. The critical temperature $T_c$ is given by the
condition $M_2(\vec P_2=0,\mu,T_c)=2\mu$, see e.g.~\cite{Alm:1994db}
for a discussion of similar aspects in nuclear matter. Implementation
of such effects into the three-body problem would require the notion
of condensate/pairing at finite $P_2$ which is a difficult problem
that has hardly been investigated.  We presently neglect this as well
as the appearance of other two-body correlations in the Fermi
functions when solving the three-body problem. However, in its turn,
the Dyson approach (cluster mean field expansion) utilized here has
the potential to re-evaluate the condition of critical temperature in
the presence of strong three-body correlations; that is, however, not
the goal of our present work.

The subtraction imposed by condition (\ref{eqn:Bound}) in the
denominator of eq.~(\ref{eqn:tau}) makes $\tau(M_2)$ finite. Note
that $\gD B(M_{2B})\rightarrow 0$ for large momenta $\vec k_\perp$, so that
$\gD B(M_{2B})$ is finite. The resulting expression for the two-body
propagator is then given by
\begin{eqnarray}
\tau(M_2) &= &\Big(i
\left[\gk(M_{2B}) {\rm arctan}2\gk^{-1}(M_{2B})\right.
\nonumber\\
&&\left.-\gk(M_{2}) {\rm arctan}2\gk^{-1}(M_{2})\right]/(2\pi)^2
\nonumber\\
&&+ \gD  B(M_{2B}) -\gD  B(M_{2})\Big)^{-1},
\label{eqn:tau2}
\end{eqnarray}
where
\begin{equation}
\gk(M_2)= \sqrt{\frac{m^2}{M_2^2}-\frac{1}{4}}.
\end{equation}
The quark mass $m(\mu,T)$ as well as the two-body masses $M_2(\mu,T)$,
$M_{2B}(\mu,T)$, and $M_{20}(\mu,T)$ depend on the chemical potential
and the temperature of the medium. Also there is an additional
dependence on the cluster's c.m. momentum related to the momentum
dependence of the blocking factors. Hence several changes of
kinematical variables are necessary in the two-body amplitude before
eq.~(\ref{eqn:tau2}) may be employed in a three-body cluster. They are
discussed in the following. The two-body mass is
\begin{eqnarray}
M_2^2&=&(P_{3} - q)^2\nonumber\\
&=&(M_3-q^+)\left(M_3-\frac{\vec q_\perp^2+m^2}{q^+}\right)-\vec q_\perp^2,
\label{eqn:M2}
\end{eqnarray}
where $q$ denotes the momentum of the odd particle.  The
three-body system is taken at rest, $P^\mu_3=(M_3,0,0,0)$. Also the
blocking dependent part $\gD B$ needs to be revisited, since the
blocking factors depend on the cluster momentum and hence $M_2$ is
different for a moving system not only because of eq.~(\ref{eqn:M2}),
but also due to medium effects that depend on the momentum. As a
consequence the arguments of the blocking factors of eq.~(\ref{eqn:B})
have to be properly replaced by
\begin{equation}
F(x,y;\vec k_\perp,\vec q_\perp)=f(x,\vec k^2_\perp)
+f(1-x-y,(\vec k + \vec q)^2_\perp).
\label{eqn:block}
\end{equation}
In the three-body rest frame $x=k^+/M_3$ and $y=q^+/M_3$.

\begin{figure}
\begin{center}
    \epsfig{figure=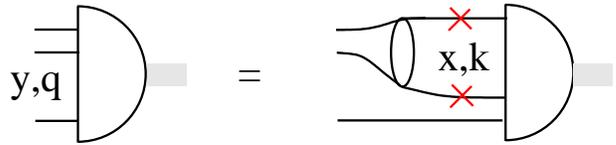,width=0.45\textwidth}
\end{center}
\caption{\label{fig:B3}
  Diagrammatic representation of the in-medium AGS-type equation for a
  zero range interaction.  The crosses refer to the Pauli-blocking
  factor $N_2$. The two-body input is given in
  Fig.~\protect\ref{fig:T2} and Eq.~\protect\ref{eqn:tau}}.
\end{figure}

The homogenous AGS-type in-medium equations for finite temperatures
and densities for the nonrelativistic bound states have been given in
Refs.~\cite{Beyer:1999zx,Beyer:2000ds}. A diagrammatic representation
of these equations for a zero range interaction is given in
Fig.~\ref{fig:B3}.  Based on the Fock space representation, a
derivation of relativistic three-body equations on the light front is
formally identical to the nonrelativistic case, if we neglect
antiparticle degrees of freedom. It appears that the modifications
required to arrive at an in-medium relativistic three-body equation
for (spin averaged) quasi particles on the light front are close to
the modifications needed in the nonrelativistic case. The reason is
the formal similarity of the light front and nonrelativistic
approaches after the approximations that have been mentioned above.
Hence a relativistic AGS-type equation on the light front including
the effects of finite temperature and density can then be written as
(compare Ref.~\cite{Frederico:1992uw,deAraujo:1995mh} for the isolated
case)
\begin{eqnarray}
\lefteqn{\Gamma(y,\vec q_\perp) = \frac{i}{(2\pi)^3}\ \tau(M_2)
\int_{M^2/M_3^2}^{1-y} \frac{dx}{x(1-y-x)}}\nonumber\\
&&\int^{k_\perp^{\mathrm{max}}} d^2k_\perp
\frac{1-F(x,y;\vec k_\perp,\vec q_\perp)}
{M^2_3 -M_{03}^2}\;\Gamma(x,\vec k_\perp),
\label{eqn:fad}
\end{eqnarray}
where we have introduced vertex functions $\Gamma$ and $\tau(M_2)$
given before. Here
\begin{equation}
k_\perp^{\mathrm{max}}=\sqrt{(1-x)(xM_3^2-m^2)},
\end{equation}
and the mass of the virtual three-particle state (in the rest system)
is
\begin{equation}
M_{03}^2=\frac{\vec k^2_\perp+m^2}{x}
+\frac{\vec q^2_\perp+m^2}{y}
+\frac{(\vec k+\vec q)^2_\perp+m^2}{1-x-y},
\end{equation}
which is the sum of the on-shell minus-components of the three
particles.

The equations suggested here, which result from a systematic Dyson
equation approach, differ formally from those of
Ref.~\cite{Pepin:2000hs}. For the case $T=0$ at finite densities the
blocking factors could be replaced by $f(\mu,T)\rightarrow
n(k_F(\mu))=\theta(k-k_F)$ where in the notation of
Ref.~\cite{Pepin:2000hs} $\mu(T=0)=\sqrt{k_F^2+m^2}$. In
Ref.~\cite{Pepin:2000hs} {\em all} three quark momenta are restricted
according to eq.~(\ref{eqn:sym})
\begin{equation}
(1-n_1)(1-n_2)(1-n_3)
\label{eqn:sym}
\end{equation}
instead of $1-n_1-n_2$ (and permutations) from eqs.~(\ref{eqn:block})
and (\ref{eqn:fad}).  Note that in these three-body equations an
additional blocking of the spectator particle does not arise in a
three-body equation driven by a two-body interaction.

\begin{figure}
\begin{center}
    \epsfig{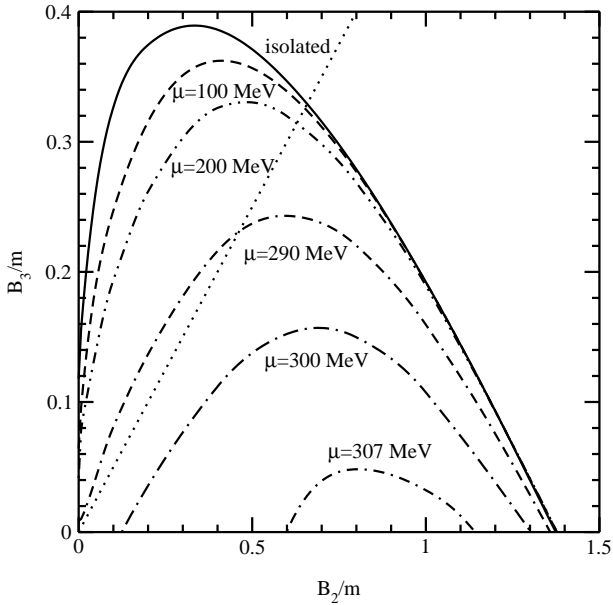}
\end{center}
\caption{\label{fig:B2toB3}
  Two-quark vs. three-quark binding energies, $B_n(\mu,T)/m(\mu,T)$,
  at $T=10$ MeV for different chemical potentials $\mu$ as indicated.
  Dotted line: $M_2/2=M_3/3$.}
\end{figure}

Presently, instead of deriving and solving two- and three-body
equations for one specific model, our intention is to explore the
correlations between two- and three-particle binding energies in a
medium of finite temperatures and densities and explore the three-body
Mott transision.  For the time being we chose values of $m(\mu,T)$
provided earlier by a Nambu-Jona-Lasinio
model~\cite{Klevansky:1992qe,Schmidt:1994di}.  That model's
approximations are close to those used in the present solution of the
relativistic three-body equation with a simple zero range interaction.
The values used for this calculation are given in
Table~\ref{tab:mass}.  Quark masses are chosen for values of $\mu$ and
$T$ to study the Mott transition.

\begin{table}[htbp]
    \caption{\label{tab:mass}
      Quark masses $m(\mu,T)$ for different temperatures $T$ and
      chemical potentials $\mu$ used in the
      calculation~\protect\cite{Schmidt:1994di}. All values in MeV.
      For the isolated case $m=300$ MeV.}
  \begin{center}
    \begin{tabular}{c|ccc}
     $\mu\backslash T$&    10 &     50  &    100\\\hline
     100 &300 &299 &279\\
     200 &300 &293 &236\\
     300 &289 &-       &-\\
     307 &279 &-       &-
    \end{tabular}
  \end{center}
    \caption{\label{tab:B3weak}
      Three-body binding energies for weak 
 coupling ($B_2/m=10^{-3}$) and $\mu=100$ MeV.}
  \begin{center}
    \begin{tabular}{crrr}
      $T$[MeV] &10 &50 &100\\\hline
      $B_3$[MeV] &23 &19 &5.0\\
      $B_3/m$[\%] &7.6 &6.5 &1.8\\
    \end{tabular}
  \end{center}
\end{table}
The correlations between the two-body and three-body binding energies,
$B_2$ and $B_3$ are shown in Fig.~\ref{fig:B2toB3} for a temperature
of $T=10$ MeV. The binding energies are defined by
\begin{eqnarray}
  B_3(\mu,T)&=&m(\mu,T)+M_{2B}(\mu,T)-M_{3B}(\mu,T)\\
  B_2(\mu,T)&=&2m(\mu,T)-M_{2B}(\mu,T).
\end{eqnarray}

Binding energies are shown in units of the respective quark masses at
the temperature and chemical potential indicated. The dotted line
indicates $M_2/2=M_3/3$ (compare~\cite{Pepin:2000hs}) and in a simple
chemical picture of an ideal gas (law of mass action) the relative
importance of the clusters. The isolated case (i.e.  no medium) is
shown as a solid line.  As the chemical potential increases (and the
quark mass becomes smaller) three-body correlation become weaker for a
given $B_2$. In this particular case $T=10$ MeV the three-body bound
state disappears for $\mu\simeq 310$ MeV, even for a model that allows
for a two-quark bound state.

In the weak coupling case $B_2\sim 0$ three-body correlations appear
stronger, however, for a given temperature the corresponding bound
states vanish above a certain value for the chemical potential. For a
given chemical potential $\mu=100$ MeV and in the weak coupling limit
the respective values for $B_3$ are given in Table \ref{tab:B3weak}.

The dependence of $B_3(T)$ for a given chemical potential is opposite
from the nuclear case, compare Ref.~\cite{Beyer:1999zx}. This is in
agreement with the expectation from other approaches that give a
negative slope for the phase transition, see
e.g.~\cite{Alford:2001dt}.  For strong couplings, i.e.  $B_2/m\gtrsim
0.65$ and $T=10$ MeV, two-body correlations dominate for all chemical
potentials.

\begin{figure}
\begin{center}
    \epsfig{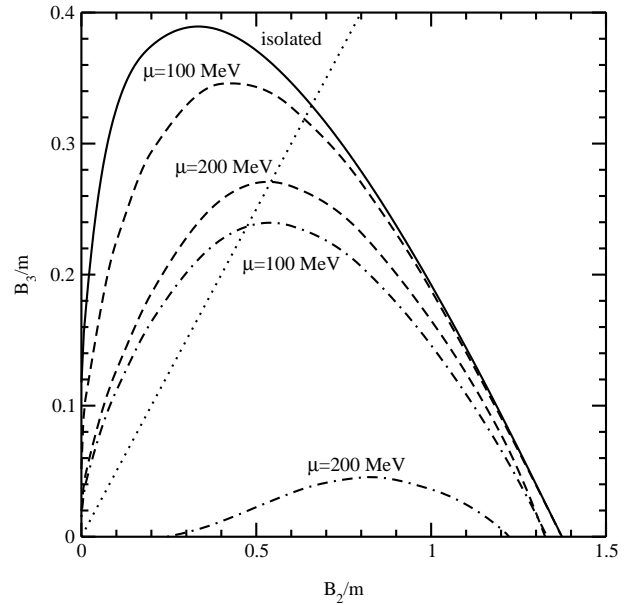}
\end{center}
\caption{\label{fig:Tmu}
  Two-quark vs. three-quark binding energies for different chemical
  potentials $\mu$ and temperatures $T=100$ MeV (dashed lines) and
  $T=50$ MeV (dashed-dotted lines). Dotted line: $M_2/2=M_3/3$.}
\end{figure}

In Fig.~\ref{fig:Tmu} the same correlations are shown for different
temperatures and chemical potentials. For higher temperatures and
chemical potentials the three-body bound state disappears and also the
three-body correlations become weaker as the lines are below the
long-dashed line. In all examples the isolated case is limiting. An
increase in $\mu$ means more particles in the medium that
preferentially occupy the momentum components necessary to form bound
states. A similar effect arises for increasing temperatures.  For a
given chemical potential $\mu=100$ MeV Fig.~\ref{fig:Tmu} reflects
stronger correlations at lower temperatures.

\begin{figure}
\begin{center}
    \epsfig{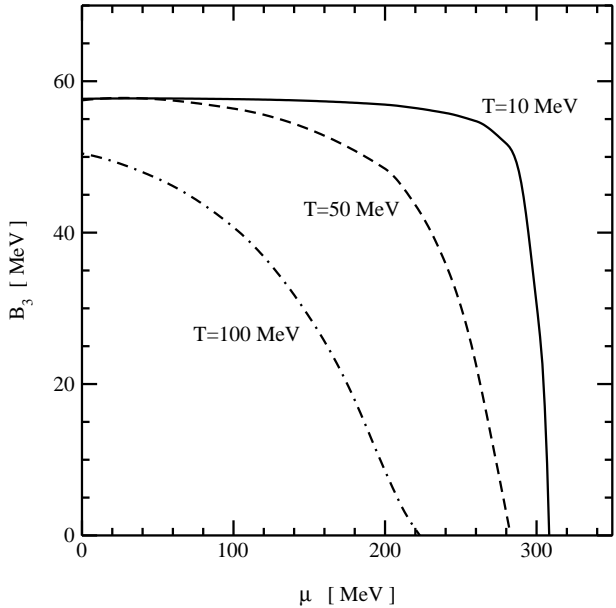}
\end{center}
\caption{\label{fig:mott}
  Binding energy of the three-body system for $M_2/m=1$ for different
  temperatures as indicated. $B_3=0$ corresponds to the Mott density.}
\end{figure}

In Fig.~\ref{fig:mott} we show the dependence of the three-body
binding energy on the chemical potential for different temperatures
$T$ for an assumed two-body bound state of $M_2/m=1$. The continuum
(for 2q+1q break-up) is reached at $B_3=0$ for certain chemical
potentials known as Mott transition. For increasing temperature the
chemical potential where the Mott transition occurs becomes smaller.

\begin{figure}
\begin{center}
    \epsfig{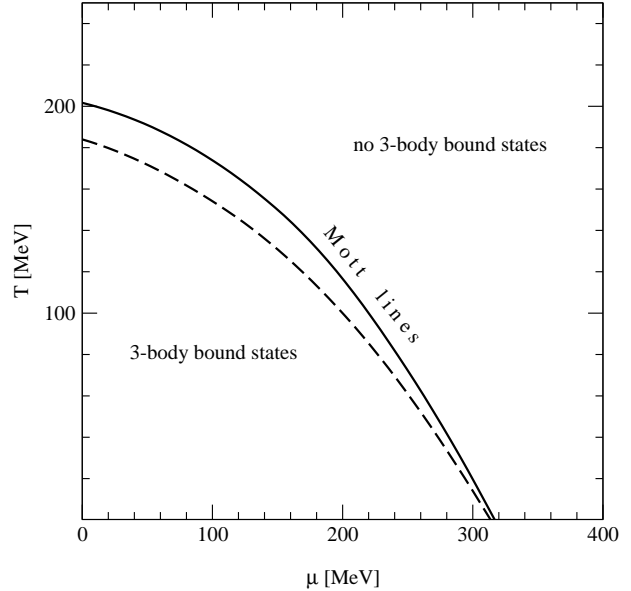}
\end{center}
\caption{\label{fig:phase}
  Mott line for the three-body system at rest in the medium with
  $M_2/m=1$ (solid) and$M_2/m=7/4$ (dashed). For values of $T$ and
  $\mu$ below the Mott lines three-body bound states can be formed.}
\end{figure}

The Mott lines, i.e. the values of $T$ and $\mu$ where the transitions
occur are given in Fig.~\ref{fig:phase} which is our main result.
This is in qualitative agreement with the confinement-deconfinement
phase transition~\cite{Alford:2001dt}.  In these cases we have assumed
a value for the two-quark bound state of $M_2/m=1$, which is close to
a value predicted by the NJL model~\cite{Pepin:2000hs} and for a
lesser bound system of $M_2/m=7/4$ to study the sensitivity on the
two-body input. It is obvious that the qualitative behavior of the
Mott transition retains for the two different models.

The onset of superfluidity is expected for $M_2(P_2=0,\mu,T_c)=2\mu$.
We mention here that for e.g. for $T_c=10$ MeV this condition is
fulfilled at $\mu= 150$ MeV for $M_2=m$ and at $\mu=260$ MeV for the
case $M_2=7m/4$.  However, no definite conclusion can be made on the
basis of the present treatment of the two-body amplitude.

In conclusion, we have derived for the first time a consistent
relativistic three-body equation for particles embedded in a medium of
both finite temperature and finite density. This equation systematically
includes the effects of Pauli blocking and mass shift of the
quasi-particles involved. The equations are solved for parameters
close to the phase transition of QCD. We find that three-body clusters
become less stable for a denser or hotter system than the two-body
cluster. This justifies, {\em a posteriori} our restriction to the
two-body bound states in this investigation. However correlations may
still exist as the ``pole'' moves into the continuum (e.g.
corresponding to anti-bound states in this approximation). 

For further investigations and to achieve more quantitative results a
specific model can be chosen that leads to a specific two-body
$t$-matrix and a corresponding three-body bound state. Furthermore,
the approximations we used in the treatment of spin of the particles
have to be relaxed and a full treatment needs to be implemented along
the lines discussed at length in Ref.~\cite{Beyer:1998xy}.  The
neglect of confinement in the vicinity of the Mott line may be
justified by color screening discussed in Ref.~\cite{Karsch:1988pv}.
However a more realistic treatment of confinement is highly
desirable~\cite{Gribov:1999ui}. The light front approach leading to
3-dim equations is suited to include relativistic confining potentials
into the formalism.

In view of the present simplifications, related to the averaging of
spin degrees of freedom and the use of a zero range force, we would
like to emphasize that the main result of this Letter is to show that
an in-medium three-body equation that reflects the basic requirements
of relativity can be derived and solved in the physical region of
interest close to the Mott transition.  The light-front framework
appears to be very useful in this context as it allows for a
Hamiltonian formalism close to an intuitive interpretation based on
nonrelativistic approaches. 

Addressing the question of color-superconductivity in a next step the
full three-body $t$-matrix has to be implemented in a calculation of
the critical temperature of condensation/pairing. This can also be
achieved in a systematic fashion within the Dyson equation approach
used here.

{\em Acknowledgment:} Work supported by the Deutsche
Forschungsgemeinschaft, grant BE 1092/10-1.

\end{document}